\documentclass[myepj-spec,pdftex]{mySvjour}
\usepackage{graphicx}
\usepackage{hyperref}
\usepackage{color}
\usepackage{xspace}
\usepackage[square,sort&compress,numbers]{natbib}   
\usepackage{verbatim}
\usepackage{amsmath,amssymb,amsfonts} 
\usepackage{tabularx}
\usepackage{multicol}
\usepackage{listings}
\usepackage[switch, modulo]{lineno}

\usepackage{lmodern,bm}                
\usepackage[T1]{sansmath} 
\SetMathAlphabet{\mathsfbf}{sans}{\sansmathencoding}{\sfdefault}{bx}{sl}
\usepackage{etoolbox}
\AtBeginEnvironment{sansmath}{}{}{}

\usepackage{lipsum}

\newcommand{\pT}{\ensuremath{p_{\mathrm{T}}}}

\newcommand{\ET}{\ensuremath{E_{\mathrm{T}}}}
\newcommand{\MET}{\mbox{\ensuremath{\not \!\! \ET}}}

\definecolor{darkblue1}{rgb}{0,0,.2}
\definecolor{darkblue}{rgb}{0,0,.2}
\definecolor{darkred}{rgb}{0.5,0,0}
\pagecolor{white} 
\color{black}     
\hypersetup{breaklinks=true, 
	colorlinks=true, 
	linkcolor=darkblue1, 
	menucolor=darkblue1, 
	urlcolor=darkblue1,
	citecolor=darkblue1,
	pdftitle={},
	pdfauthor={},
	pdfsubject={},
	pdfkeywords={},
	pdfproducer={}
}
\usepackage{siunitx}

\bibstyle{plain}
\newcommand{\bi}{\begin{itemize}}
\newcommand{\ei}{\end{itemize}}

\newcommand{\ben}{\begin{enumerate}}
\newcommand{\een}{\end{enumerate}} 

\newcommand{\bt}[1]{\begin{table}[tb]\begin{tabular}{#1} \hline\hline  \\[-1.0em]}
\newcommand{\et}[2]{\hline\hline \end{tabular} \caption{#1} \label{#2} \end{table}}

\newcommand{\be}{\begin{equation}}
\newcommand{\ee}{\end{equation}}

\newcommand{\bea}{\begin{eqnarray}}
\newcommand{\eea}{\end{eqnarray}}

\newcommand{\bc}{}

\newcommand{\mev}{\ensuremath{\mathrm{\,Me\kern -0.1em V}}\xspace}
\newcommand{\gev}{\ensuremath{\mathrm{\,Ge\kern -0.1em V}}\xspace}

\begin{document}
	
	\twocolumn[{%
		\begin{@twocolumnfalse}
			
			\begin{flushright}
				\normalsize
			\end{flushright}
			
			\vspace{-2cm}
			
			\title{\Large\boldmath Studies on the $H\rightarrow bb$ cross section measurement at the LHeC with a full detector simulation}
			%

\author{Subhasish Behera \inst{1} \and Bernard Brickwedde \inst{1} \and Matthias Schott \inst{1}}
\institute{\inst{1} Institute of Physics, Johannes Gutenberg University, Mainz, Germany}

			
			\abstract{
The future Large Hadron electron Collider (LHeC) would allow collisions of an intense electron beam with protons or heavy ions at the High Luminosity–Large Hadron Collider (HL-LHC). Owing to a center of mass energy greater than a TeV and very high luminosity ($\sim$1 $ab^{-1}$), the LHeC would not only be a new generation collider for deep-inelastic scattering (DIS) but also an important facility for precision Higgs physics, complementary to $pp$ and $e^+e^-$ colliders. Previously, it has been found that uncertainties of $0.8\%$ and $7.4\%$ can be achieved on the Higgs boson coupling strength to $b$- and $c$-quarks respectively. These results were obtained in the fast simulation frameworks for the LHeC detector. Focusing on the dominant Higgs boson decay channel, $H\rightarrow bb$, the present work aims to give a comparison of these results with a fully simulated detector. We present our results in this study using the publicly available ATLAS software infrastructure. Based on state-of-the art reconstruction algorithms, a novel analysis of the $bb$ decay could be performed leading to an independent verification of the existing results to an exceptionally high precision.}	
	\maketitle
	\end{@twocolumnfalse}
}]

\tableofcontents
	
\section{Introduction}	
A future Large Hadron Electron Collider (LHeC) \cite{AbelleiraFernandez:2012cc, Agostini:2020fmq} at CERN would collide $7$ TeV LHC proton beams and $60$ GeV electron beams at a luminosity of $10^{34}~\text{cm}^{-2}\text{ s}^{-1}$. This would take place in parallel to the proton-proton collisions of the LHC. The design for this electron accelerator is based on a linac-ring $ep$ collider configuration with two superconducting linacs, each below 1km in length, operating in continuous wave (CW) energy recovery mode \cite{Agostini:2020fmq, Bartnik:2020pos}. The main focus of the LHeC physics program is deep inelastic scattering (DIS) physics, probing a completely new area of the low-$x$ phase space, allowing the precise determination of proton and nuclear parton distribution functions (PDFs) \cite{Lai:2010vv}. PDFs are an essential pre-requirement for any fututre high energy hadron collider(s), including LHC. In addition to the DIS program, searches for physics beyond the Standard Model, such as leptoquarks, contact interactions, as well as RPV and SUSY promise significantly higher sensitivities than is currently possible at existing colliders. The LHeC also has significant potential for measurements in the Higgs sector. Here, Higgs boson production via vector-boson-fusion and its decay into $b$- and $c$- quarks could be much cleaner than at the LHC \cite{PhysRevD.82.016009}, allowing us to probe the relevant Higgs couplings to a higher precision. This is due to the clean final state, absence of pile-up, unique and simple Higgs production mechanism, and the redundant reconstruction of the DIS kinematics using both the leptonic and the hadronic final state. Therefore, it is of great interest to study the prospects for Higgs production in $ep$ collisions and examine the possible decay modes carefully. The previous results, published in Refs.~\cite{AbelleiraFernandez:2012cc, Agostini:2020fmq, PhysRevD.82.016009}, have been obtained in several independent analyses. They used a \textsc{Delphes} \cite{deFavereau:2013fsa} based simulation framework adapted especially to the DIS environment of the LHeC detector.\par
The current analysis aims to independently verify the previous results using the full event simulation framework of the ATLAS detector which appears to be well suited within the limitations of detector differences to such a comparison. The current study proceeds in three steps: firstly, we reproduce the results of previous studies based on the \textsc{Delphes} LHeC detector simulation using a cut-based approach for the signal selection. Secondly, we repeat the same study using the fast simulation, also based on the \textsc{Delphes} framework of the ATLAS detector, which has a slightly different acceptance and detector response functions. Finally, we repeat the study using the official full simulation and reconstruction infrastructure of the ATLAS experiment and compare the results with those of the fast simulations. This three-step procedure will permit an evaluation of the relevance and validity of certain assumptions in such analyses. It should be noted that, in Ref.~\cite{Agostini:2020fmq} a dedicated effort was made to refine the final result through careful acceptance and background studies using a \textit{Boosted Decision Tree} (BDT) algorithm to obtain a result, which was extended to the six dominant Higgs decay channels. The focus here is on a cut based comparison of $H\rightarrow bb$ analysis approaches, to look for principal possible differences for which the reaction was simulated.\par
The paper is structured as follows: The differences between a future LHeC detector and the ATLAS Experiment at the LHC are briefly discussed in \autoref{sec:ComparisonLHeCATLAS}. Details of the Monte Carlo (MC) event generation and detector simulation are discussed in \autoref{sec:McSimulation}. Our implementation of the previous analysis approach \cite{Agostini:2020fmq} is validated for the study of $H\rightarrow bb$ at the LHeC and the transfer of this to the ATLAS experiment, based on fast detector simulations, is discussed in \autoref{sec:PreviousAnalysis}. The results based on the full detector simulations are summarised and discussed in \autoref{sec:Full}. A short discussion on forward electron tagging and its impact on backgrounds are discussed in \autoref{sec:ForwardElec}. Finally, we conclude our discussion in \autoref{Sec:Conclusion}.

\section{Comparison of a dedicated LHeC detector and the ATLAS experiment\label{sec:ComparisonLHeCATLAS}}

The proposed LHeC detector design has to maximise the coverage in the forward and backward regions of the colliding beam, and be asymmetric in beam direction. This reflects the asymmetry in the energy of the colliding particles \cite{Agostini:2020fmq}. The detector dimensions are of the order of $13$m in length and $9$m in diameter, allowing the reuse of the magnet from the L3 experiment. Hence, this experiment has a much smaller footprint than that of the ATLAS and CMS detectors.\footnote{In the following we use a right-handed coordinate system to describe the ATLAS and LHeC detectors with its origin at the nominal interaction point (IP) and the positive $z$-axis along the proton beam direction. The $x$-axis points from the IP to the center of the LHC ring, and the $y$-axis points upward. Cylindrical coordinates $(r,~\phi)$ are used in the transverse plane, $\phi$ being the azimuthal angle around the $z$-axis. The pseudo-rapidity is defined in terms of the polar angle $\theta$ as $\eta = -\ln \tan(\theta/2)$.  Angular distance is measured in units of $\Delta R \equiv \sqrt{(\Delta\eta)^{2} + (\Delta\phi)^{2}}$}\par
Starting from the beam line and moving outward, the inner most component is a tracking detector for the reconstruction of charged particles with a transverse momentum resolutions down to $10^{-3}$ GeV$^{-1}$ with an impact parameter resolution of $10~\mu$m. The coverage in pseudo-rapidity for the inner barrel is $|\eta|<3.3$, larger than the current setup of the ATLAS detector, which allows a track reconstruction within $|\eta|<2.5$.\footnote{It should be noted that the upgrade of the ATLAS detector for the high luminosity phase of the LHC foresees an extension of the tracker coverage} The inner detector is followed by an electromagnetic (EM) calorimeter which could be based on liquid argon technology, similar to ATLAS. The hadronic calorimeter of an LHeC detector could be based on an iron-scintillator setup, surrounds the $3.5$T magnet coil system and is enclosed within a muon tracker system. The coverage of the full calorimeter system is the same as the tracking detector, i.e. $-4.3<\eta<4.9$ which is comparable to the ATLAS coverage of $|\eta|<4.9$. The muon system of an LHeC detector and  that of ATLAS cover a pseudo-rapidity range up to $4.0$ and $2.7$ respectively. However, this difference is of minor importance for this study. One of the largest differences between ATLAS and LHeC detectors are dedicated calorimeters in the end caps of the LHeC detector to precisely measure forward high energy products (silicon-tungsten) or the scattered electron (silicon-copper). The identification and reconstruction of electrons in the forward region is of particular important to classify neutral current processes.\par
In order to study $H\rightarrow bb$ processes, the identification of particle jets that originate from $b$-quarks, known as $b$-tagging, is typically based on information of secondary vertices and track impact parameters, i.e. observables which are based on the inner detector of the experiment. Given the significantly larger coverage of the tracking detector of the LHeC compared to ATLAS, the $b$-tagging coverage is expected to be significantly improved. However, the minimal transverse momentum requirement on particle jets in the $H\rightarrow bb$ process is $20$ GeV for the cut-based analysis, implying nearly no jets originating from $b$-quarks beyond $\eta>3.0$, see \autoref{sec:PreviousAnalysis} below. The effective difference in the $b$-tagging between an LHeC detector and ATLAS is therefore rather small. We argue that the effective coverage of the ATLAS detector is therefore comparable to the acceptance of the LHeC detector when it comes to the study of the $H\rightarrow bb$ and $H\rightarrow cc$ processes. 	

\begin{figure*}[htb]
\begin{center}
\begin{minipage}{0.23\textwidth}
\resizebox{1.0\textwidth}{!}{\includegraphics{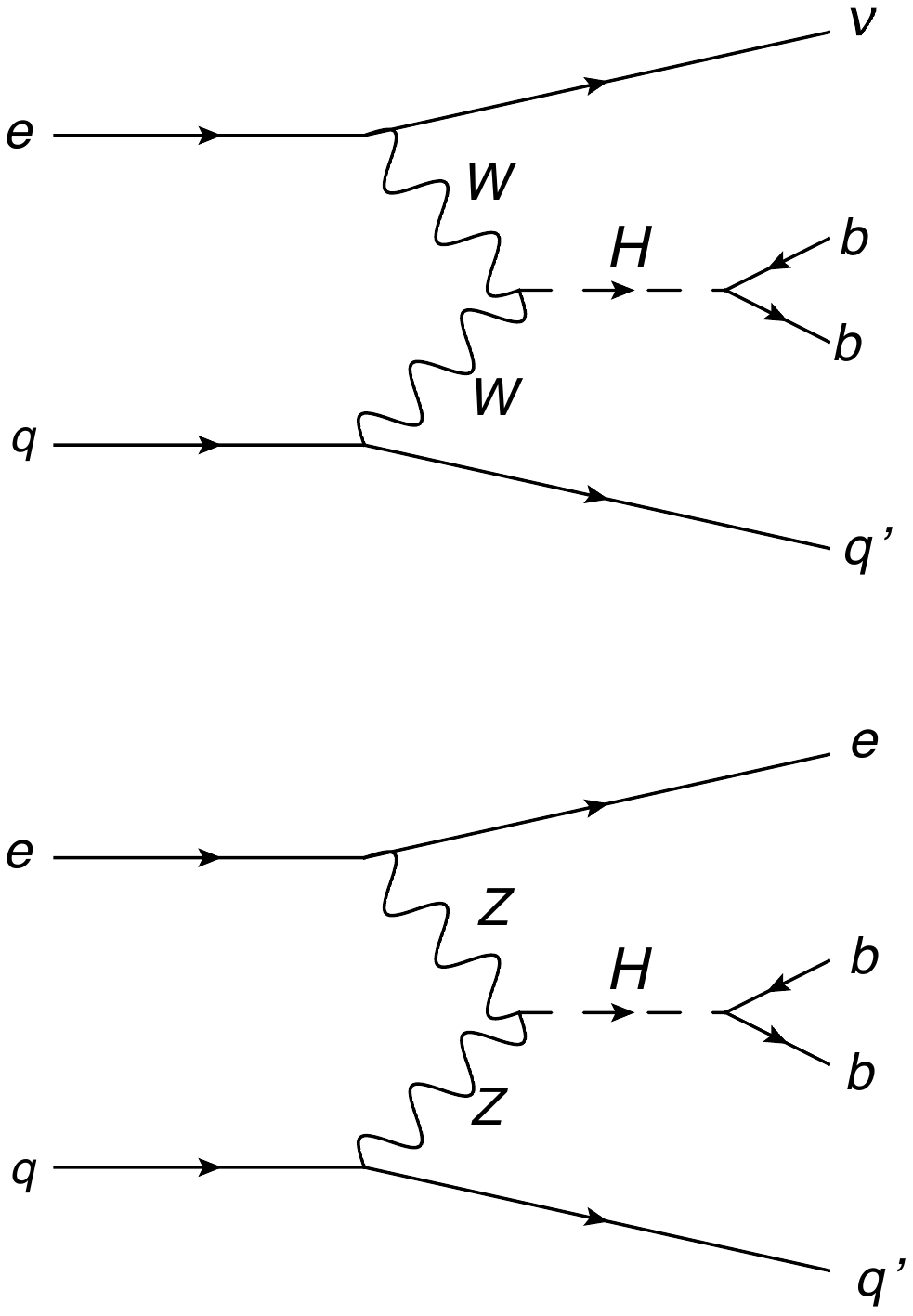}}
\caption{\label{fig:Signal}Leading Order Feynman diagrams for the production of Higgs bosons at the LHeC.}
\end{minipage}
\hspace{0.3cm}
\begin{minipage}{0.70\textwidth}
\resizebox{1.0\textwidth}{!}{\includegraphics{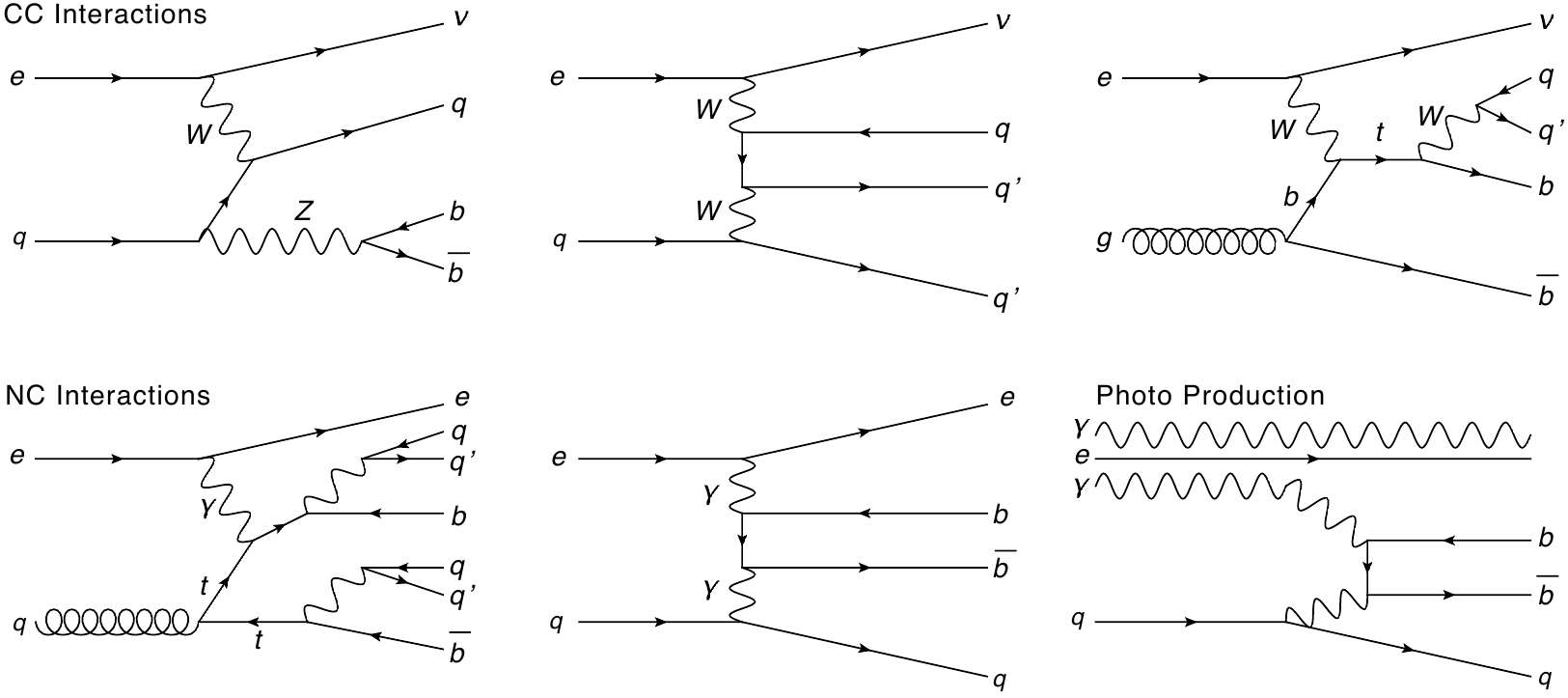}}
\caption{\label{fig:Background}Example Feynman diagrams for background processes. Upper row: charged current interactions, Lower row: neutral current interactions and photo-production. \vspace{0.25cm}}
\end{minipage}
\end{center}
\end{figure*}

\section{Monte Carlo samples and detector simulation \label{sec:McSimulation}}

The proposed baseline energy of the electron beam is $60$ GeV, which in combination with $7$ TeV proton beam results in a center of mass energy of $1.3$ TeV. This is about four times that of its predecessor, HERA~\cite{Wiik:1985sb} at DESY which had a center of mass energy of $319$ GeV. In addition the expected luminosity is about three orders of magnitude higher.\par
Higgs bosons in $ep$ collisions will be produced through vector boson fusion via either a charged current (CC) or a neutral current (NC) interaction, depicted in \autoref{fig:Signal}. Since the production cross section for charged current interactions is dominant, we focus only on the associated final state of $\nu_e b\bar b j$ in this work, where $j$ represents the jet of the scattered quark of all flavours except for top flavour. The most relevant background processes are shown schematically in \autoref{fig:Background} and can be distinguished between CC and NC induced processes: (i) CC multi-jets with no intermediate $Z$ bosons, top quarks and Higgs boson, (ii) CC single top, (iii) CC on-shell $Z$ boson decay to two $b$-quarks, (iv) photo-production of multi-jet final states, (v) photo-production with a $t\bar t$ final state, (vi) NC on-shell $Z$ boson decay to two $b$-quarks, (vii) NC with $e^-b\bar b j$ final state and (viii) NC with $e^-t\bar t$. The top quark further decays to jets in the final state within the SM. The CC and NC processes can be distinguished by the appearance of an electron in the final state of a NC interaction.\par
The \textsc{MadGraph5} generator \cite{Alwall:2014hca} has been developed to model the hard scattering of proton-electron collisions for all relevant signal and background processes, using the CTEQ6L1 PDF set \cite{Pumplin:2002vw}. The factorisation and renormalisation scales are taken as the mass of Higgs boson, $m_h=125$ GeV for the signal processes, while a dynamical scale setup has been used for the background processes. The showering and hadronisation of the hard scattering events was carried out using \textsc{Pythia8.303} \cite{Sjostrand:2014zea}. To control the cross section of the background processes during the event generation, several requirements on the transverse momentum, $\pT$, pseudo-rapidity, $\eta$, of the final charged leptons and quarks as well as on the invariant mass of two final state quarks have been applied.\par
At least 100k events for all signal and background samples have been produced, in order to get sufficient statistics after final selection. A summary of all generated samples, including the applied generator-level cuts and the corresponding cross section predictions, are summarised in \autoref{tab:event_sample}. 

\begin{table*}[h]
\scriptsize
\begin{tabularx}{\linewidth}{l | l  | l | c | c | c }
\hline
Short 			&	Process								& Generator-Level Cuts	 			& Generator			& Cross Section	& Number Of	\\
Description		&										& 					 			& 					& [pb]			&  Events$\times10^3$	\\
\hline
Signal			&	$p e^- \to \nu_{e} h q \to   \nu_{e} b \bar b q$		& $|\eta| <10,~m_{jj}>60$				& MadGraph5+		  	&0.09997			&150			\\	
				&										&								&Pythia8					&				&			\\
\hline
CC-qqq			&	$p e^- \to \nu_\ell q q q $					& $\pT^q>10~GeV,~\pT^b>10~GeV,$	& MadGraph5+				&5.49074			&214			\\
				& (w.o. $Z$, H, $t$, $\bar t$ )										& $ |\eta| <10,~m_{jj}>70,~m_{bb}>70$	&	Pythia8			  	&				& 			\\
\hline
CC-top			&	$p e^- \to \nu_\ell \bar t q$					& $\pT^q>10~GeV,~\pT^b>10~GeV,$	& MadGraph5+	&0.36820			&137		\\
				&										& $ |\eta| <10,~m_{jj}>70,~m_{bb}>70$	& 	Pythia8			 	&				& 			\\
\hline
CC-Z			&	$p e^- \to z q \nu_\ell,  \to b \bar b q \nu_\ell $	& $\pT^q>10~GeV,~\pT^b>12~GeV,$	& MadGraph5+ 	&0.13631			&107	\\
				&										& $ |\eta^q| <5.5,~|\eta^b| <5.5,~|\eta^\gamma| <5,$	& 	Pythia8			 	&				& 			\\
				&										& $|\eta^\ell| <5,~m_{jj}>60,~m_{bb}>60$	&					&				&			\\
\hline
PA bbq			&	$p \gamma \to b \bar b q$  				& $\pT^q>12~GeV,~\pT^b>19~GeV,$	& MadGraph5+	& 0.90876			&100		\\
				&										& $ |\eta^q| <5.2,~|\eta^b| <3.5,~|\eta^\gamma| <10,$	& 	Pythia8			&				& 			\\
				&										& $|\eta^\ell| <10,~m_{jj}>115,~m_{bb}>115$	& 				&				& 			\\
\hline
PA-tt				&	$p \gamma \to t \bar t \to bqq\bar bqq	$	& $\pT^q>10~GeV,~\pT^b>12~GeV, $	& MadGraph5+	&0.00876			&100		\\
				&										& $|\eta^q| <5.5,~|\eta^b| <4,~|\eta^\gamma| <10,$	&  Pythia8			&				& 			\\
				&										& $|\eta^\ell| <10,~m_{jj}>80,~m_{bb}>80$& 				 	&				& 			\\
\hline
NC-Z			&	$p e^- \to Z q e^- \to b \bar b q e^-$			& $\pT^q>10~GeV,~\pT^b>12~GeV,$		& MadGraph5+  	&0.02246			&100		\\
				&										& $\pT(\ell)>0.01~GeV,~ |\eta^q| <5.5,~|\eta^b| <5.5,$	& 	Pythia8		&				& 			\\
				&										& $|\eta^\gamma| <10,~4<|\eta^\ell| <10,~m_{jj}>60,~m_{bb}>60$	& 	  	&				& 			\\
\hline
NC-bbq			&	$p e^- \to e^- b \bar b q $					&$\pT^q>10~GeV,\pT^b>12~GeV,$		& MadGraph5+   	& 2.37302			&100		\\
				&										&$\pT(\ell)>0.01~GeV, |\eta^q| <5.5,~|\eta^b| <4,$	&	Pythia8		&				& 			\\
				&										&$|\eta^\gamma| <10,~4<|\eta^\ell| <10,~m_{jj}>80,~m_{bb}>80$		&   	&				& 			\\
\hline
NC-tt				&	$p e^- \to e^- t \bar t \to e^- b q q \bar b q q$		& $\pT^q>10~GeV,~\pT^b>12~GeV,$	& MadGraph5+	&0.81091			&100		\\
				&										& $\pT(\ell)>0.01~GeV,~ |\eta^q| <5.5,~|\eta^b| <4,$& 	Pythia8		  	&				& 			\\
				&										& $|\eta^\gamma| <10,~4<|\eta^\ell| <10,~m_{jj}>80,~m_{bb}>80$& 	  	&				& 			\\
\hline
\end{tabularx}
\caption{The cross section of the signal and all possible background samples for corresponding generator-level cuts are shown in the table.\protect\footnotemark~Here, $q$ represents either a quark or anti-quark of any SM flavor except top quark and $\ell = (e^\pm, \mu^\pm)$. Whenever we mention $b$-quark (anti-$b$-quark) we specify that the generator has at least the same number of $b$-quarks (anti $b$-quarks) at the parton level.}\label{tab:event_sample}
\end{table*}

Once all sample for the signal and background processes are available on generator level, the detector response for a future LHeC detector and ATLAS has been simulated. Two different approaches have been used here: the \textsc{Delphes} framework~\cite{deFavereau:2013fsa} allows for a fast simulation of an approximated detector response for typical detector in high energy particle physics. The simulation includes a tracking system within a magnetic field, electromagnetic (ECAL) and hadronic calorimeter (HCAL) as well as a muon system. High-level objects like isolated electrons, particle jets or missing transverse energy are reconstructed using observables such as tracks and energy deposits in the calorimeter. The stable charged particles on generator level with a minimal transverse momentum (e.g. $p_\text{T}>100$\,MeV) are subjected to track reconstruction. The track reconstruction efficiency as well as the resolution and the momentum scale is parameterised against $p_\text{T}$, $\eta$ and $\phi$. Particles on generator level that reach the calorimeter system deposit energy in the electromagnetic and hadronic calorimeter cells. The relevant cells can then be grouped together in one calorimeter tower, which are then used for jet reconstruction as well as the calculation of missing transverse energy. The resolutions of the electromagnetic and hadronic calorimeters are independently parameterised depending on the particle kinematics, a stochastic term, a noise term, and a constant term. Reconstruction and identification efficiencies of leptons are also parameterised within the \textsc{Delphes} software. The resulting reconstructed objects, e.g. particle-jets, electron, or muons which are used for actual physics analysis, only provide a first approximation of a real detector response. Similar to the previous physics studies for the LHeC, we use \textsc{Delphes} for the fast simulation of the LHeC detector response, based on a dedicated configuration file \cite{LHeCDelphesCard}. In addition, we use the standard ATLAS configuration file available in the \textsc{Delphes} software framework, based on Refs.~\cite{ATLAS:2015dex, Kulchitsky:2000gg, ATLASElectromagneticBarrelCalorimeter:2006ymk}, for the fast simulation of the ATLAS detector.\par
In contrast to a fast detector simulation, a \textit{full} simulation of a particle collisions in an LHC detector starts from the theoretical modeling of the interaction (event generation), resulting in particles which can be considered stable during their passage through the actual detector. The interaction between these particles and the detector are typically simulated using the \textsc{Geant4} framework \cite{Agostinelli:2002hh}, which is able to simulate the interaction between final-state particles and the detector on the microscopic level. In the \textsc{Geant4} simulation, each particle produced by the event generator is tracked step-by-step through the simulated detector. At each step, physical processes such as decays and interactions with material are simulated. If the interaction takes place in an active part of the detector, a \textit{hit} is recorded. From these hits, the simulated response of the sub-detector is calculated in a process called digitisation. The output of this process forms raw data objects (RDOs) which should be of the same format as the real detector electronics are expected to deliver. Based on these RDOs, dedicated reconstruction algorithms are applied, which inference all relevant physical observables, such as the momentum, the trajectory, charge and flavor of particles. Therefore, it results in significantly more realistic predictions in particular when it comes to the reconstruction of fake signatures, e.g. a reconstruction of an electron which actually was caused by a particle jet. \footnotetext{The signal cross-section is corrected for the correct branching ratio of $H\rightarrow bb$.}\par 
The ATLAS software framework, Athena \cite{Aad:2010ah}, which is based on the Gaudi framework \cite{Barrand:2001ny}, contains the full simulation workflow of the experiment, starting from event generation, simulation and digitisation, up to reconstruction. It is publicly available \cite{ATLASSofteware} and was setup on the Mainz computing cluster Mogon, independently from the ATLAS Collaboration. It was used to fully simulate all relevant signal and background samples in \autoref{tab:event_sample}. In a second step, we convert the event information at reconstruction level of the full simulation to the same format used for the studies based on the fast simulation samples.\par
While the effect of triggers at the LHeC or ATLAS was not studied here explicitly, we argue that a dedicated $H\rightarrow bb$ trigger will have a sufficiently high efficiency so that the impact on the subsequent analysis is minimal.

\section{Validation of the analysis strategy using the fast-simulation of a LHeC detector \label{sec:PreviousAnalysis}}

The full potential of Higgs physics at the LHeC can only be realised using advanced analysis techniques, as is discussed in detail in \cite{Agostini:2020fmq}. However, we need a baseline analysis model which allow for direct and simple comparison of the detector response for several different signal and background processes. We argue that a validation of a simple cut-based analysis with a full simulation consolidates the more advanced techniques. In a first step, the cut-based LHeC  $H\rightarrow bb$ analysis \cite{Agostini:2020fmq, Ellis:Thesis} has been repeated using the \textsc{Delphes} LHeC detector simulation on our samples. Jets, reconstructed with {\it anti-kT} algorithm, with cone size of R = $0.4$ and a minimal transverse momentum of $\pT>5$ GeV within a pseudo-rapidity range of $|\eta|<6.0$ are pre-selected. Events with a reconstructed electron in the forward region are vetoed in order to suppress NC interactions. For the LHeC study it is assumed that an additional forward electron tagging will be available which efficiently reduces NC processes as well as the final state signatures of photo-production \cite{Agostini:2020fmq}. We therefore do not consider the $p e^- \to e^- b \bar b q $ and  $p e^- \to e^- t \bar t \to e^- b q q \bar b q q$ process for the validation of our results.\par
All remaining events are required to pass several kinematic selection requirements to select DIS induced processes: The missing transverse energy $\MET$, defined as the negative vector sum of all reconstructed cluster energies in the transverse plane, is required to be greater than $30$ GeV. Moreover, the fraction of the electron energy carried by the (virtual) propagator in the proton rest frame, $y_h$, calculated using the \textit{Jacquet-Blondel} method \cite{Amaldi:1979qp}, where $y_h=\sum_{hadrons} \frac{E-p_z}{2E_e}$ with $E_e$=60 GeV, is required to be smaller than $0.9$. In addition, the negative transferred four momentum squared, $Q_h^2=\frac{\MET^2}{1-y_h}$, has to be larger than $500$ GeV$^2$.\par
Since the signal process yields two $b$-quarks and one light-quark in the final state, each event is required to contain at least three reconstructed particle jets with a $\pT>20$ GeV. Two of these jets must be $b$-tagged, i.e. identified to be originated from a $b$-quark, within the detector region $|\eta|<2.5$. The jet with the highest $\pT$ which is not $b$-tagged is referred as light-jet throughout the following passage. The top-quark related background processes are vetoed by excluding events with an invariant mass of the two $b$-jets and the light-jet below $250$ GeV and events for which the invariant mass of one $b$-jet and the light-jet is below $130$ GeV.\par
\begin{figure*}[tb]
\centering
\includegraphics[trim=0 -17 0 0,clip,width=0.48\linewidth]{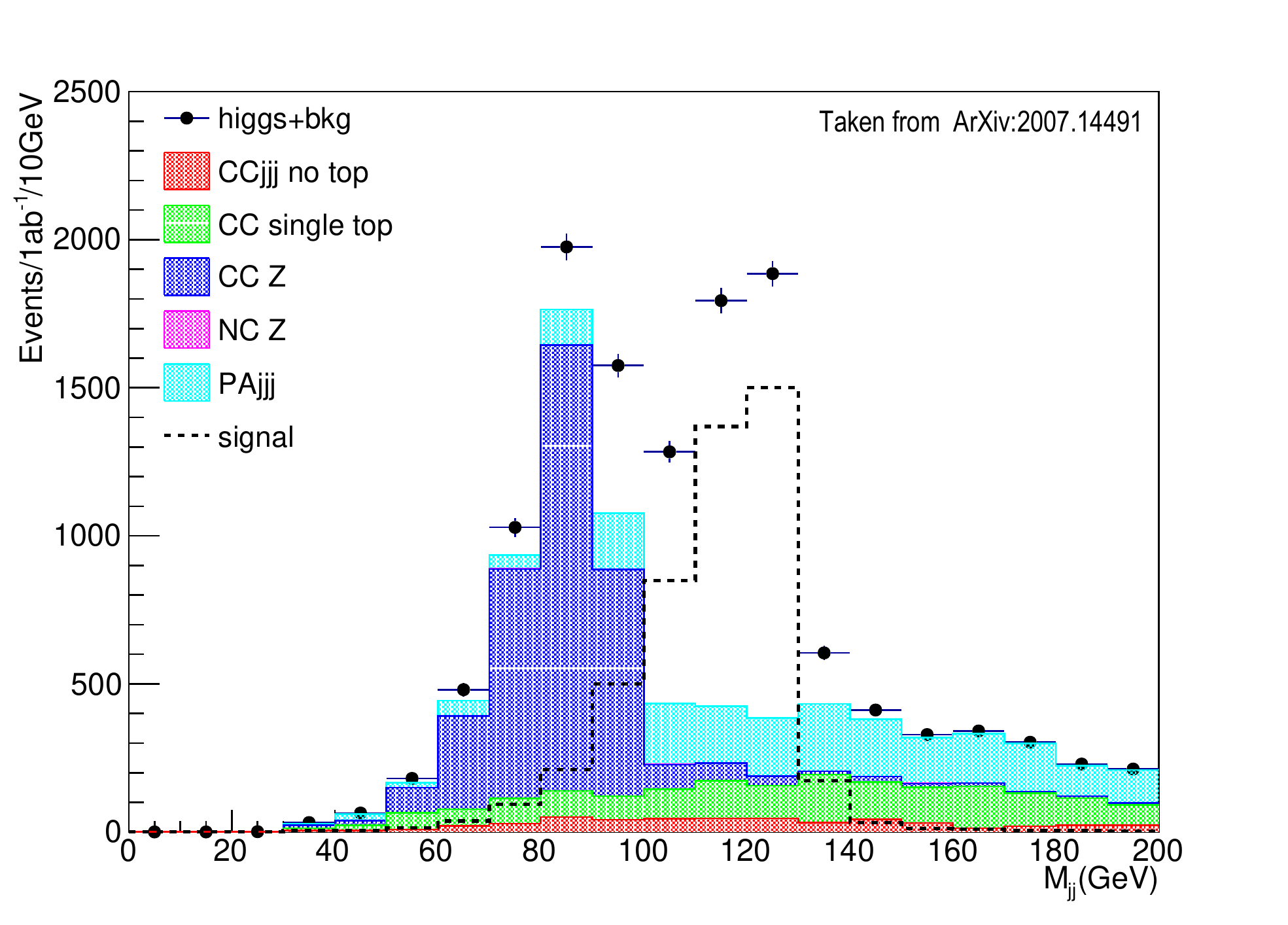}
\includegraphics[trim=0 0 0 0,clip,width=0.48\linewidth]{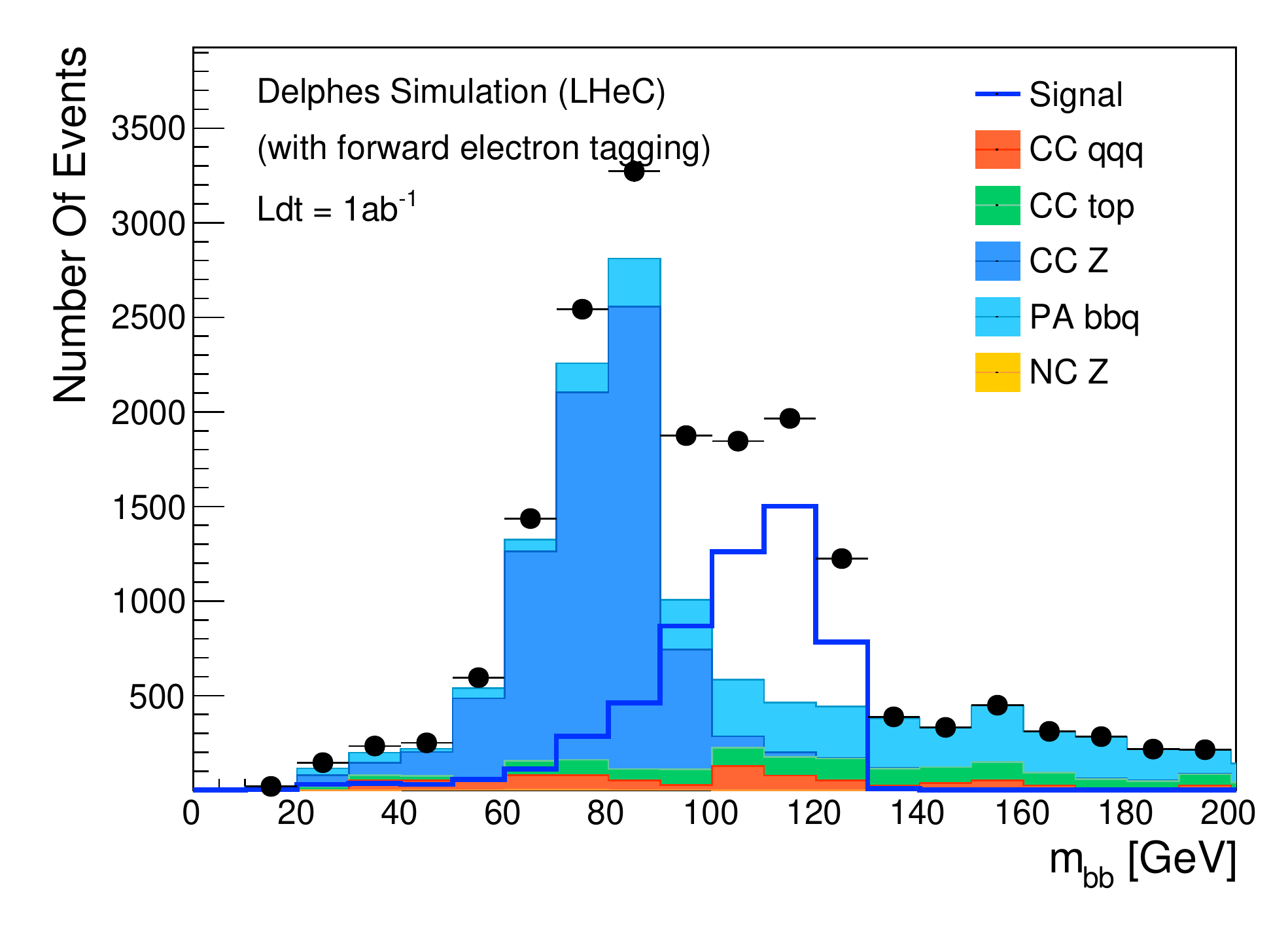}
\caption{The Invariant di-jet mass of two selected $b$-tagged jets of the updated LHeC CDR \cite{Agostini:2020fmq} (left) and the analysis from our study (right).}
\label{fig:LHeCResults}
\end{figure*}

Furthermore, the events are required to have at least one reconstructed jet in the forward region, where $\eta>2.0$, and the $\Delta\Phi$ values between the $b$-jets and $\MET$ is required to be larger than $0.2$. The invariant mass of the two $b$-jets, $m_{bb}$, is required to be within a window of 100 and 130 GeV, which is defining the final signal region. The expected event yield as well as the $m_{bb}$ distribution for the signal and background processes, for an integrated luminosity of ${\it \int L dt} = 1 ab^{-1}$, after the event selection are shown in \autoref{tab:eventyieldsLHeC} and \autoref{fig:LHeCResults}. These include both the cut-based LHeC CDR analysis and our analysis. 

\begin{table}[tph]
\centering
\resizebox{0.98\linewidth}{!}{
\begin{tabular}{l | c | c} \hline
Process		& LHeC CDR (Delphes)	& This Study (Delphes) \\\hline
Signal             & 3720 	& 3540$\pm$50  \\\hline
CC-qqq    		& 157		& 200$\pm$70  \\
CC-top    		 & 339		& 310$\pm$30  \\
CC-Z    		  & 173		  & 90$\pm$10    \\
PA bbq			& 606		& 840$\pm$90  \\\hline
\end{tabular}}
\caption{Expected event yields for signal and background processes in the signal region ($100<m_{bb}<130$), for ${\it \int L dt}= 1~ab^{-1}$, from the cut-based analysis of the official LHeC CDR \cite{Agostini:2020fmq} (taken from Figure \ref{fig:LHeCResults}) alongside this study.\label{tab:eventyieldsLHeC}}
\end{table}

The expected signal over background ratio using the LHeC fast simulation changes from $2.9$ (in LHeC CDR) to $2.5\pm0.2$ (in our study) for the cut based analysis of the samples. The expected event yields in the signal region agree well for most processes within the statistical uncertainty. Several differences can be explained by different generator settings, e.g. the usage of \textsc{Pythia8} instead of \textsc{Pythia6} \cite{Sjostrand:2006za} for the showering  in this study.\footnote{We compared the effect of \textsc{Pythia6} on the showering for the signal and background samples. In this case we found comparable results.} We observe a significant difference in the predicted charged current processes involving $Z$ bosons that decay into $b$-quarks.\par
While the expected number of those background events is smaller in the signal region (i.e., for 100$<m_{bb}<$130 GeV) in our study, the overall number of CC-Z events is smaller by a factor of roughly two\footnote{This difference is due to differences in the underlying cross-section prediction of the MC Event generators used. We verified our number by the usage of alternative generators.}. However, this difference does not impact significantly the further analysis, as it contributes less than $10\%$ to the overall background contribution.\par
It should be noted that the estimated photo-production background in our study does not rely on any electron tagging in the forward region. The cross section calculation using LHeC CDR for multijet photo-production background ($p\gamma \rightarrow qqq$) is estimated to be $\sim170$~pb with a reduced invariant mass cut on two light- or $b$-quarks (i.e., by considering, min. $m_{bb}=m_{jj}=65$ GeV).\par
In order to optimise the MC production, we demand at least two $b$-jets at the generator level (for the process, $p\gamma \rightarrow bbq$) which reduces the cross-section to $\sim0.9$ pb, results in \autoref{tab:event_sample}. \autoref{fig:LHeCResults} also indicates that the shapes of the signal and backgrounds processes in the $m_{bb}$ distribution can, in general, be successfully reproduced. We observe a shift towards lower masses in the signal samples when comparing these results to the LHeC CDR, which was traced to differences in the underlying rapidity distribution of the Higgs-Boson \footnote{A possible cause could be differences in the assumed initial energy of the incoming electrons}. However, these differences will not be relevant for the purpose of our study, i.e. the validation of expected physics performance using a full detector simulation.

\section{Electron-proton collisions at the ATLAS detector using fast and a full detector simulation \label{sec:Full}}

The same signal selection cuts are applied to the analysis using ATLAS detector simulations for both the fast and full detector modules. The remaining number of the signal events after each cut is shown for three independent simulations in \autoref{tab:cutflow_1}. The largest difference is seen for the jet requirement cuts, where $25-35\%$ less events survive for the ATLAS detector, mainly due to the lower $\eta$ coverage of the detector components and differences in the $b$-tagging efficiencies for different $\eta$ values of the particle jets. However, this difference is largely mitigated by the subsequent rejection cuts for top-quark events, where relatively more events with $b$-jets at a large rapidity fail the selection. The expected signal yield between the fast simulation of ATLAS and LHeC agree within $15\%$.

\begin{table}[tph]
\centering
\resizebox{0.98\linewidth}{!}{
\begin{tabular}{l |c |c |c} \hline
Signal						  & LHeC			& ATLAS			& ATLAS 	\\
Selection					& (Delphes)		 & (Delphes)		& (Full Sim.)\\\hline
All Events  				&99970 			  & 99970  		&  99970  \\
No Electron 			  &97100 			 & 99941  	  &  98311 \\
Kinematic Cuts 			&70356 			   & 70971		&  65236\\
($MET,Y,Q^2$)		   &	 				  & 				&  		  \\
Jet Requirements	  &18325 			 & 13373  	   &  11537 \\
Top Rejection 			&6809  				& 6003   	   &  5222  \\
Forward Jet 			 &6745  			 & 5878   		&  5117  \\
$\Delta\Phi(B_{1/2}, MET)>0.2$	&5438  				& 4662   			&  3866  \\\hline
$100<m_{bb}<130$					&3540  				& 3160   			&  2270  \\\hline
\end{tabular}
}
\caption{Cutflow for the signal samples $pe\rightarrow \nu H(\rightarrow bb)j$ normalised to ${\it \int L dt}= 1~ab^{-1}$ for different detector simulations using \textsc{Delphes} for the LHeC detector and ATLAS, as well as a full simulation of the ATLAS detector.}	\label{tab:cutflow_1}
\end{table}

In a second step, the signal selection has been applied on the fully simulated signal and background samples of the ATLAS detector. The largest difference of about $10\%$ compared to the fast simulation is induced by kinematic requirements on $\MET$, $Y$ and $Q^2$, where the dominant effect arises from the $\MET$ distribution. The $\MET$ resolution is significantly worse in the full simulation compared to the assumptions made within \textsc{Delphes}, thus significantly less events pass the $\MET$ requirements. The cut-flow between the fast and full simulation is consistent until the requirement of a light jet, where a large difference of $20\%$ has been observed. A further significant difference is introduced by the requirement on the invariant mass of the two $b$-tagged jets, yielding a final difference of $30\%$. This is caused by the inferior jet energy resolution in the full simulation compared to the fast simulation. This causes a broadening of Higgs signal in the full simulation.\par
\begin{figure*}[tb]
\centering
	\resizebox{0.49\textwidth}{!}{\includegraphics{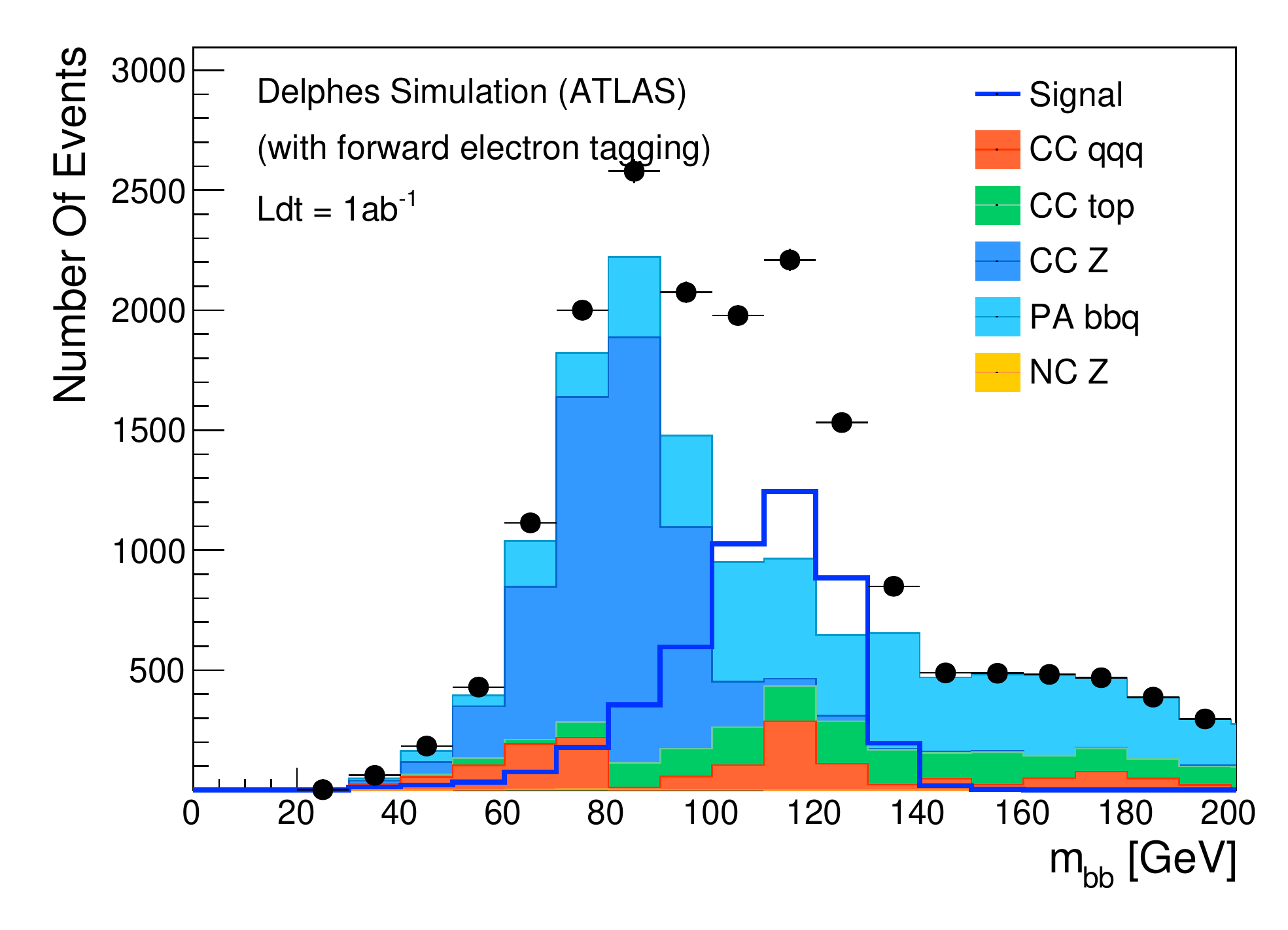}}
	\resizebox{0.49\textwidth}{!}{\includegraphics{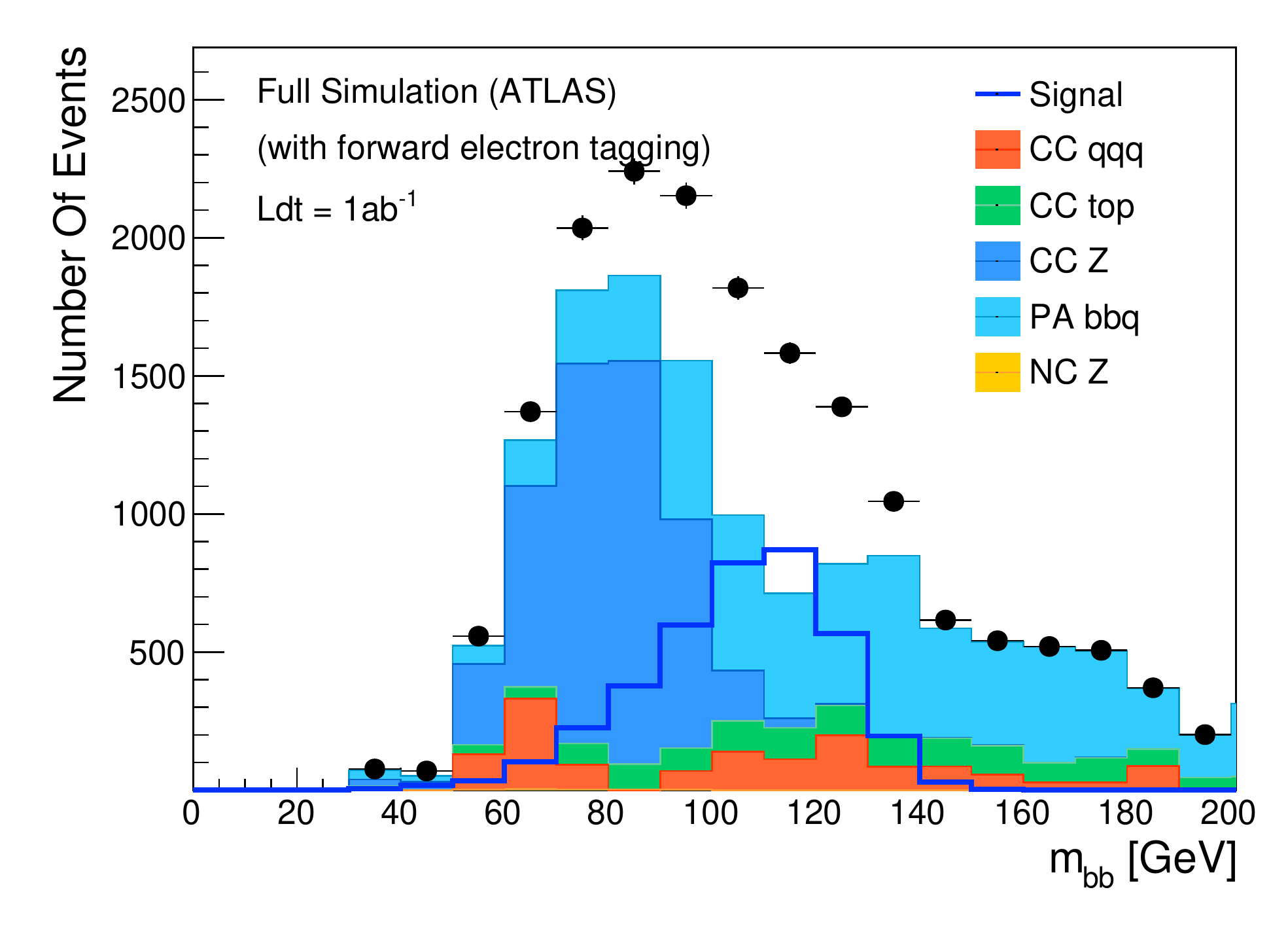}}
	\resizebox{0.49\textwidth}{!}{\includegraphics{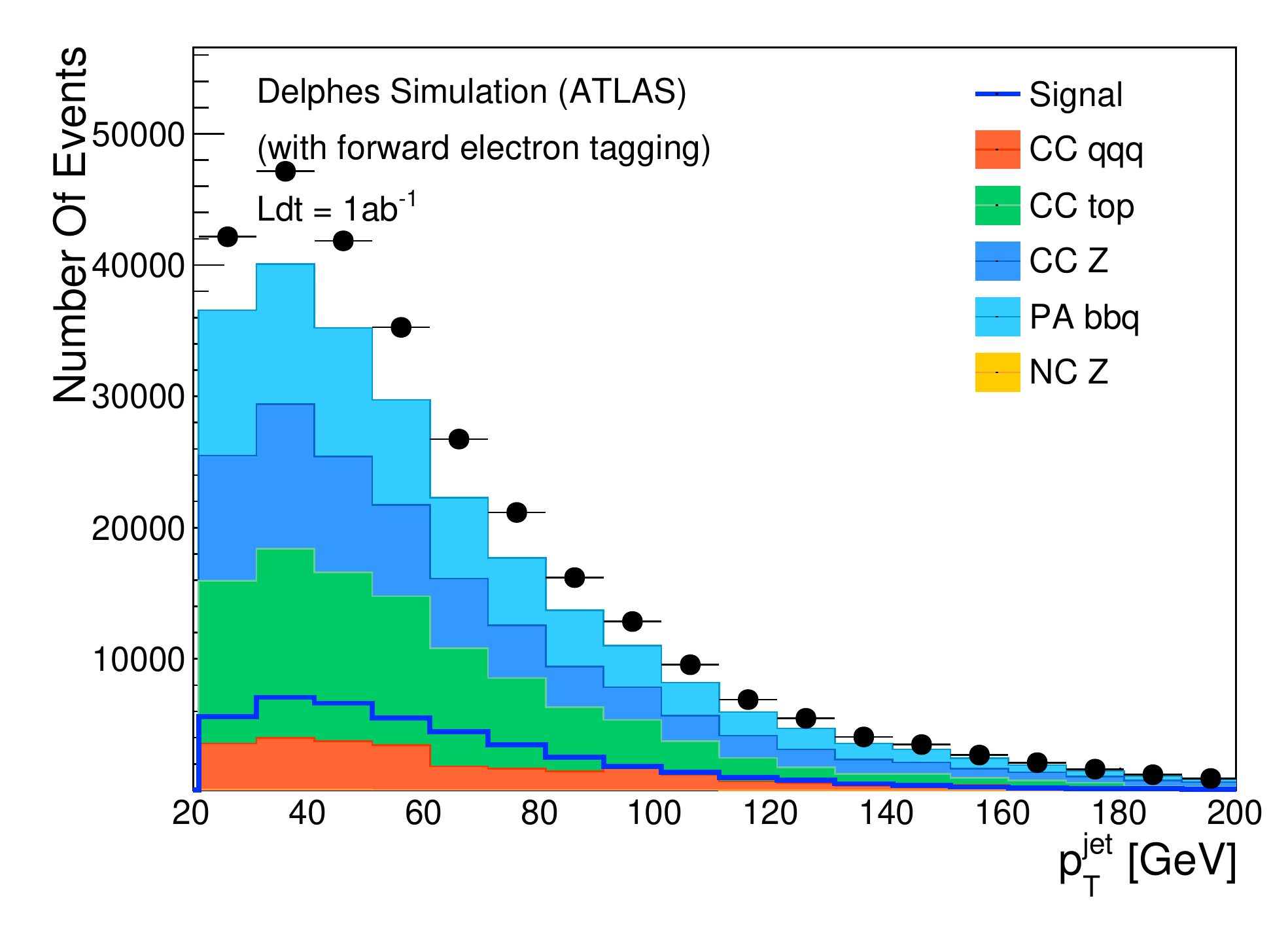}}
	\resizebox{0.49\textwidth}{!}{\includegraphics{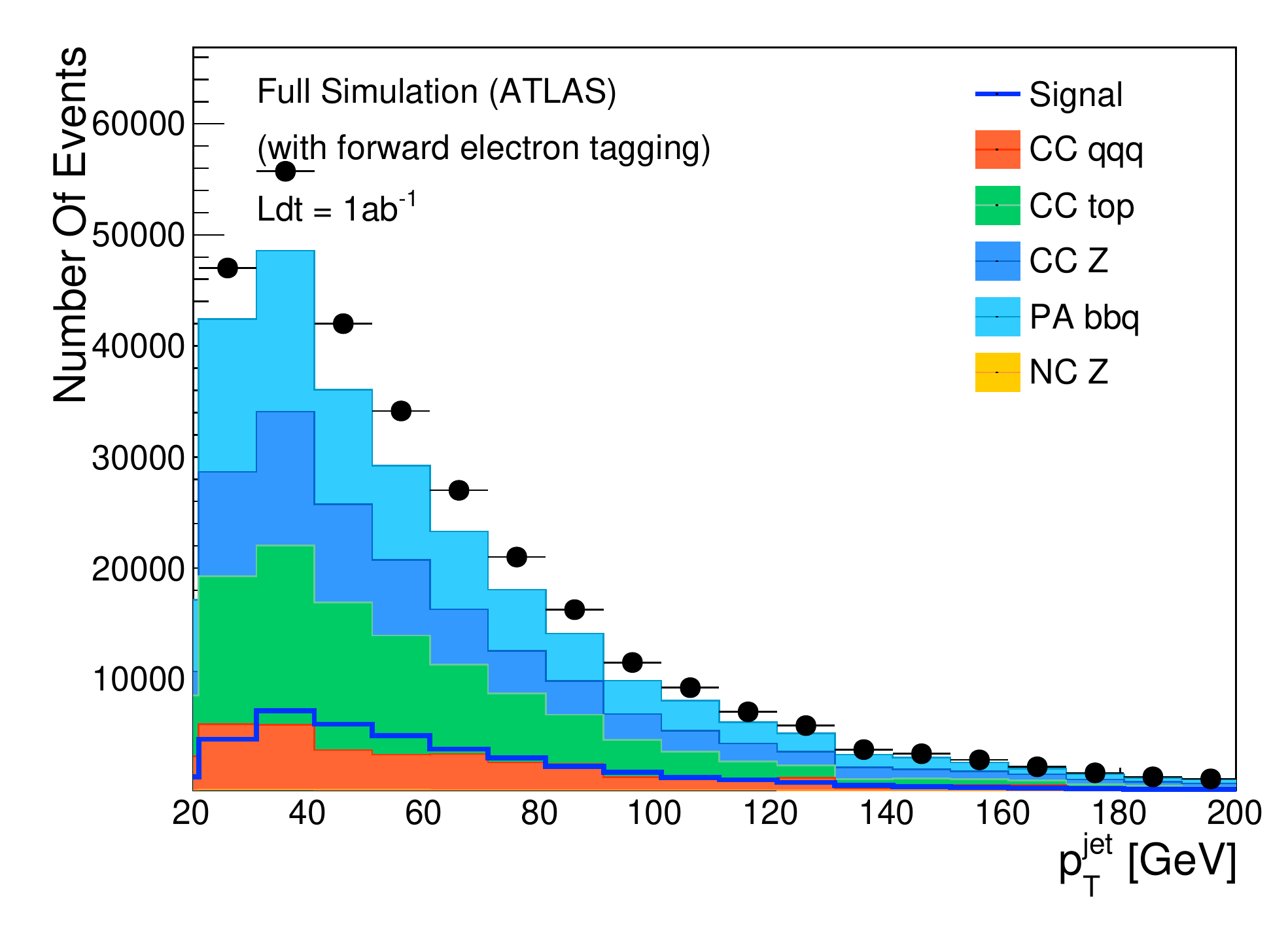}}
\caption{Invariant di-jet mass distributions of two selected $b$-tagged jets (upper row) and the $\pT$ distribution of all selected jets, i.e. two $b$-tagged and one light-flavor jet (lower row). In both cases these are performed once using the fast (left) and once the full simulation (right) of the ATLAS detector.}
\label{fig:LHeCResults_1}
\end{figure*}

The differences between the fast and full simulation for all background samples is summarized in \autoref{tab:eventyieldsLHeC_1}. Overall a good agreement can be seen. A comparison of selected kinematic distributions, namely the invariant mass of the two $b$-jets as well as the $\pT$ distribution of all selected jets for the signal and background samples using the fast and the full simulation is shown in \autoref{fig:LHeCResults_1}.

\begin{table}[tph]
\centering
\resizebox{0.98\linewidth}{!}{
\begin{tabular}{l | c | c} \hline
Process		& ATLAS (Delphes)	& ATLAS (Full Simulation) \\\hline
Signal             & 3160$\pm50$ 	 & 2270$\pm40$  \\\hline
CC-qqq    		& 500$\pm$100	  & 450$\pm$100  \\
CC-top    		& 480$\pm$40	   & 330$\pm$30  \\
CC-Z    		 & 240$\pm$20		& 220$\pm$20   \\
PA bbq		    & 1340$\pm$110	   & 1520$\pm$130 \\\hline
NC-Z    		 & 1$\pm$ 1		        & 0 $\pm$0 \\
NC-bbq 		   & 710 $\pm$130	   & 840$\pm$160   \\
NC-tt       	  & 1200$\pm$100	& 1160$\pm$110 \\
\hline
\end{tabular}}
\caption{Expected event yields for the signal and background processes in the signal region ($100<m_{bb}<130$), for ${\it \int L dt}= 1~ab^{-1}$, for both the ATLAS fast and full simulation.\label{tab:eventyieldsLHeC_1}}
\end{table}

The expected signal over background ratio using the ATLAS fast simulation is $3160/2560\approx 1.2$, while the full simulation yields $2270/2520\approx 0.9$. This difference impacts the expected precision of the cross section of the $H\rightarrow bb$ process, which can be experimentally determined via
\[
\sigma_{(H\rightarrow bb)} = \frac{N_{Data} - N_{Background}}{\epsilon \cdot {\it \int L dt}},
\]
where $N_{Data}$ and $N_{Background}$ denote the expected number of data and background events, $\epsilon$ the acceptance and selection efficiency of the signal process, and $\int L dt$ the expected integrated luminosity.\par
The statistical uncertainties on the data and the background for the different scenarios are summarised in \autoref{tab:uncertainty}. The systematic uncertainties on the selection efficiency are expected to be of a similar size to those in recent studies of top-quark pair production at the LHC \cite{ATLAS:2016oxs} and are assumed to be $\approx0.7\%$. The systematic uncertainties on the background contributions are assumed to be $2\%$, since all background processes can, in principle, be studied in dedicated control regions and hence the full theoretical uncertainty on the background prediction need not be applied. This results in overall uncertainties, of $2.5\%$ for the LHeC scenario and $3.3\%$ and $4.4\%$ for the fast and the full simulation of ATLAS detector respectively. This indicates a difference of only $1\%$ between the fast and full ATLAS simulation.

\begin{table}[h]
\centering
\resizebox{0.98\linewidth}{!}{
\begin{tabular}{l |c |c |c} \hline  
Uncertainty                 &LHeC	   &ATLAS (Delphes)	     &ATLAS (Full Sim.)\\\hline
Statistical (Data)          		  &2.0\%	 &2.4\%               &3.0\%\\
Statistical (Background)    	&1.1\%      &1.6\%                &2.2\%\\
Background Systematic	      &0.8\%     &1.6\%                &2.2\%  \\
B-Tagging and Jet-Related    &0.7\%     &0.7\%                &0.7\%\\\hline
Total                                     &2.5\%     &3.3\%                &4.4\%\\\hline
 \end{tabular}
}
\caption{Calculation of the expected uncertainties on the cross-section for a given number of signal and background event in the different scenarios.} \label{tab:uncertainty}
\end{table}

While the signal over background ratio is smaller in the full simulation mainly due to the limited jet energy resolutions, implying larger uncertainties, the general validity of the previously reported physics performance for the $H\rightarrow bb$ process is confirmed. A cross-section uncertainty on the $H\rightarrow bb$ process is expected to be on the percent level with an integrated luminosity of $\int L = 1 $ab$^{-1}$. This assumes that no further background processes contribute and NC induced interactions can be efficiently vetoed by a forward electron tagging system. However, the signal selection is not currently optimised and only a cut based approach has been implemented for the ATLAS based result. In particular, making use of advanced signal classifiers such as boosted decision tree or deep neural networks have the potential to increase the signal selection efficiency by a factor of three to six, while reducing the background contribution.\footnote{Studies using a machine learning approach on the fully simulated samples are ongoing, but are beyond the scope of this paper.} It has been suggested in \cite{Agostini:2020fmq} that the precision on the cross section can be significantly improved below $1\%$, since most experimental uncertainties, such as $b$-tagging efficiencies can be measured with high precision in data, in particular using $Z\rightarrow bb$ decays.

\begin{figure*}[thb]
\centering
	\resizebox{0.32\textwidth}{!}{\includegraphics{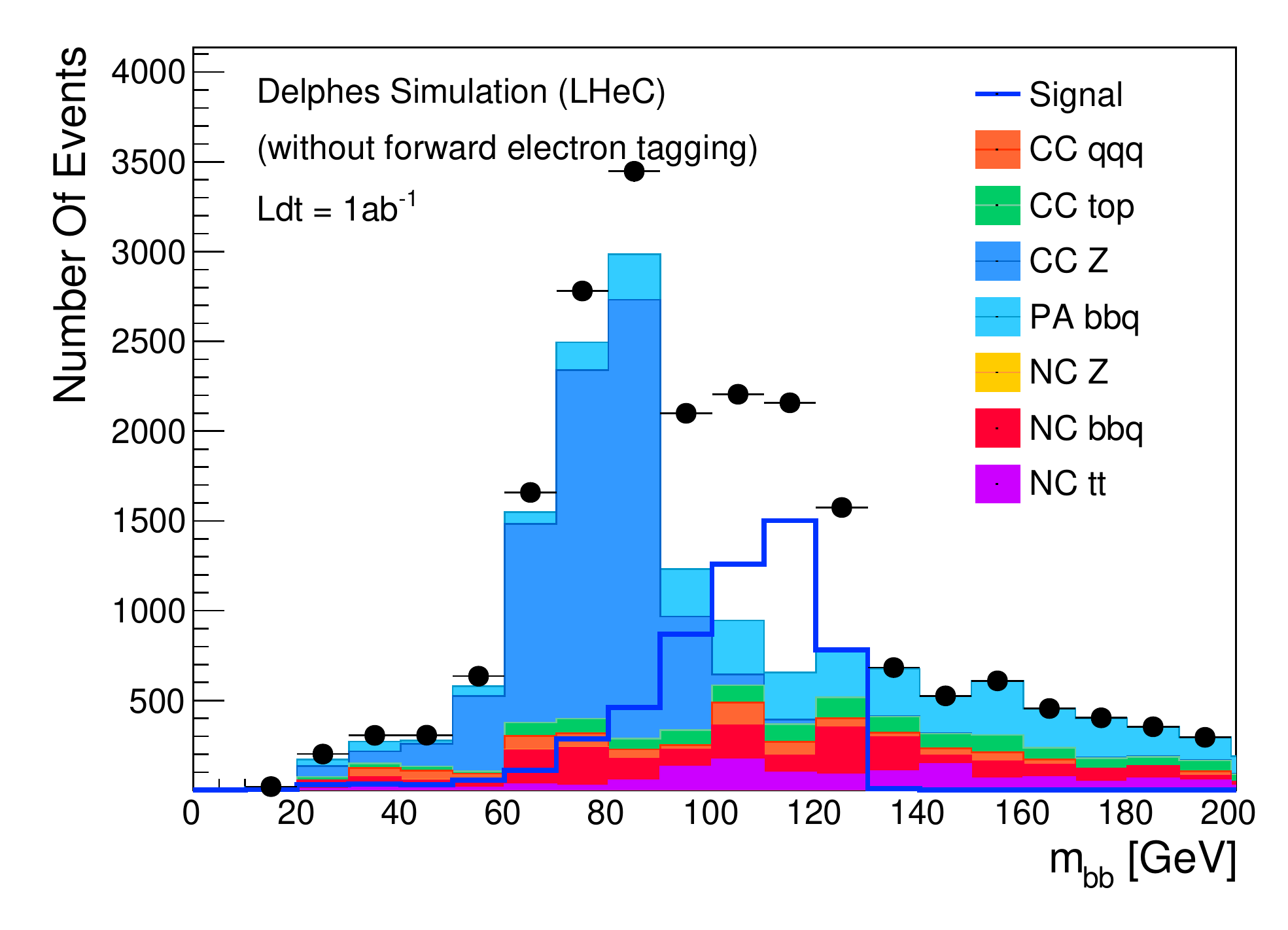}}
	\resizebox{0.32\textwidth}{!}{\includegraphics{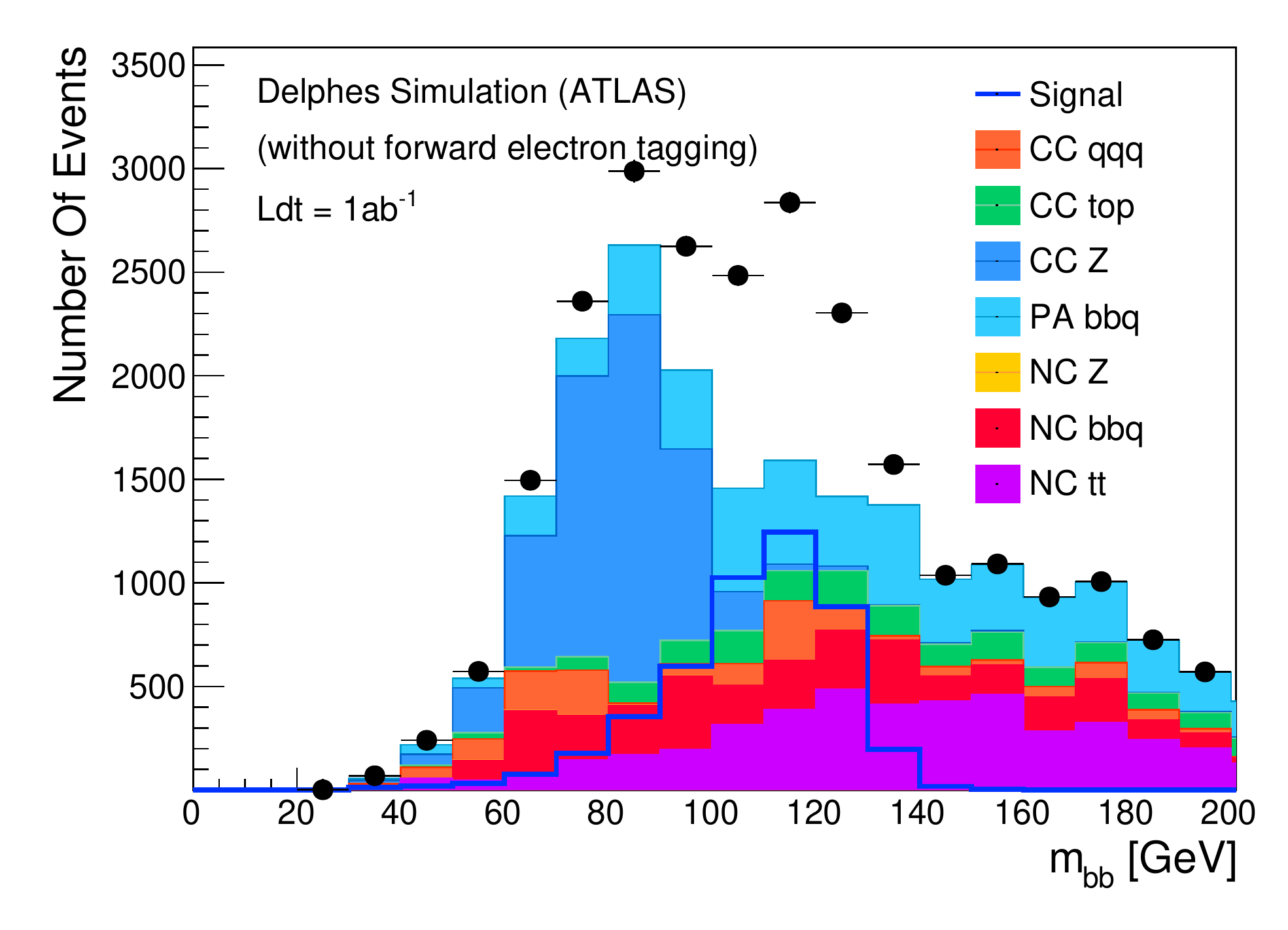}}
	\resizebox{0.32\textwidth}{!}{\includegraphics{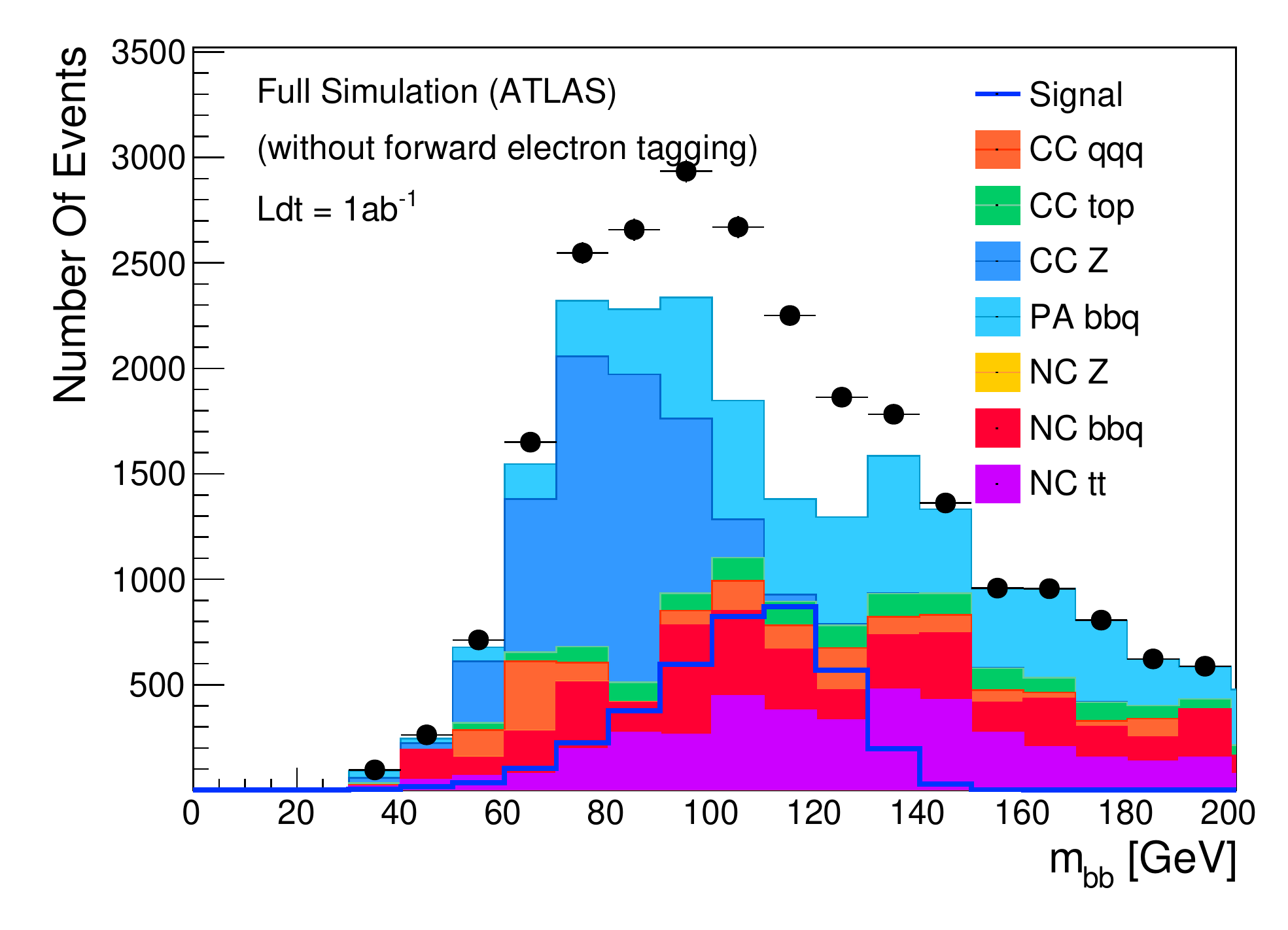}}
\caption{Invariant di-jet mass of two selected $b$-tagged jets for the signal as well as the CC, NC and photo-production background processes for the LHeC detector (left), the fast simulated ATLAS detector (middle) and the fully simulated ATLAS detector (right).}
\label{fig:LHeCResultsNotag}
\end{figure*}

Given the small kinematic differences within the fiducial phase-space definition of the $H\rightarrow bb$ study, it is valid to assume that the observed differences between the fast and the full ATLAS detector simulation will be good first-order approximation for the expected differences between a fast and full simulation of an LHeC detector with state-of-the art reconstruction algorithms. It should be noted that this only holds in the context of the $H\rightarrow bb$ process and will not be naively transferable to other processes, such as the study of DIS since here the forward detectors play a significantly larger role. Nevertheless, our study using a full detector simulation gives no indication that the expected physics performance in the Higgs boson sector at a future LHeC detector is unrealistic. 

\section{Background expectations without forward electron tagging \label{sec:ForwardElec}}

The expected rapidity range of the scattered electrons in photo-production processes that lead to a signal in the final state configuration is between $-15$ and $-5$. However, the corresponding cross sections are sufficiently small that no forward electron tagging is necessary. The scattered electron in the neutral current interactions exhibits a very low $\pT<0.5$ GeV (for more than $90\%$ of the events) and are expected to cover a rapidity range down to $-10<\eta<-4$, where the NC top-quark pair production is expected to peak at $-10$. It was so far assumed that a forward electron tagger could reject those processes. Since no simulation for such forward electron tagging exists in the \textsc{Delphes} framework, nor in any available \textsc{Geant4} based simulation, we also studied the expected background contributions when no forward electron tagging can be applied. The expected number of additional background events is also shown in \autoref{tab:eventyieldsLHeC_1}. The absence of forward electron tagging would therefore enhance the number of expected background events by nearly double the value compared with forward electron tagging, for both the fast and the full simulation. 
The resulting distributions for the signal and background processes for an LHeC detector and the fast and fully simulated ATLAS detector is shown in \autoref{fig:LHeCResultsNotag}.\par Significantly less NC top-quark pair events are expected at the LHeC, due to significantly larger coverage of the electron identification at the LHeC detector. The top-quark pair background at an ATLAS-type detector can be reduced by more than a factor of $3$, by employing the forward electron reconstruction, which is now possible within the available detector design.\par
As a preliminary conclusion, the measurement of the $H\rightarrow bb$ cross section will also be possible when no forward electron tagging is applied, but with a reduced precision. An independent cross check of the expected contributions from photo-production processes is necessary.

\section{Conclusion\label{Sec:Conclusion}}
	
In this work, we estimated the prospects of the $H\rightarrow bb$ cross section measurement at the LHeC with an integrated luminosity of ${\it \int L dt} = 1 ab^{-1}$ using the full detector simulation and state of the art reconstruction algorithms of the ATLAS Experiment. A signal over background ratio of $0.9$ and a cross-section uncertainty below $4.5\%$ are expected, where approximate statistical and systematic uncertainties have been considered. The signal selection efficiency is lower by $20\%$ in the full detector simulation, which can be explained by the differences in the jet-energy and $\MET$ resolutions.
In order to reach a sub-percent precision, the signal over background ratio would need to improve by a factor of $5$ to $6$. Given that simple optimisations on the top-quark rejection cut as well as the signal region definition in the full simulation would lead to an improvement by a factor of $2.5$, it is realistic that a multivariate analysis could yield a final signal over background ratio of the required value. Overall, our result is in agreement with the previously obtained result in a cut based analysis. However, the expected background of photo-production processes might require an additional cross check. In summary, our studies further consolidates the strong case for the LHeC as an excellent opportunity for precision studies within the Higgs sector. 

\section*{Acknowledgement}
We thank the ATLAS Collaboration for their agreement to use the ATLAS Software Framework for this work, as well as Simone Amoroso for his help in setting up the framework at the MOGON Cluster. Moreover, we would like to thank Uta Klein, who developed the original LHeC analysis and gave us important feedback on several theoretical aspects as well as highly valuable input for this paper draft. We would also like to thank Ellis Kay and Masahiro Kuze for their advise and clarifications on the MC sample production. Finally, we would like to thank Max Klein and Philip Kennedy for a final paper review. This work is supported by the German Research Foundation DFG and the grant (SCHO 1527/6-1). 
	
\bibliographystyle{apsrev4-1} 
\bibliography{./Bibliography}
	
\end{document}